\pdfoutput=1
\documentclass[fleqn,usenatbib,usedcolumn]{mnras}

\usepackage{graphicx,amssymb,hyperref}
\usepackage{multirow}
\usepackage{color}

\usepackage[fleqn]{amsmath}
\usepackage{abbrev}
\usepackage{siunitx}
\usepackage{comment} 
\usepackage{gensymb}
\usepackage{booktabs,tabularx}
\usepackage{flushend}
\usepackage{xspace}
\usepackage{physics}
 \usepackage{xcolor}
\usepackage{textcomp}
\usepackage{nicefrac}
\usepackage{units}
\usepackage{verbatim}
\usepackage[shortlabels]{enumitem}
\usepackage{subcaption}
\setlist[itemize]{leftmargin=5pt}

\def\mnras{MNRAS}
\def\apj{ApJ}

\def\simlt{\lower.5ex\hbox{$\; \buildrel < \over \sim \;$}}
\def\simgt{\lower.5ex\hbox{$\; \buildrel > \over \sim \;$}}
\def\etal{{\it et al.}}

\def\kpc{\mathrm{\, kpc}}

\def\mpc{\mathrm{\, Mpc}}
\def\msun{\mathrm{\, M_\odot}}

\def\effein{\theta_\mathrm{E,eff}}

\defcitealias{2012ApJ...761....1W}{W12}
\defcitealias{2019MNRAS.488.3646R}{R19}

\newcommand{\bahamas}{\textsc{bahamas}\xspace}

\def\gs{\mathrel{\raise1.16pt\hbox{$>$}\kern-7.0pt \lower3.06pt\hbox{{$\scriptstyle \sim$}}}}         
\def\ls{\mathrel{\raise1.16pt\hbox{$<$}\kern-7.0pt \lower3.06pt\hbox{{$\scriptstyle \sim$}}}}   

\newcommand{\vect}[1]{\boldsymbol{#1}}

\newcommand{\be}{\begin{equation}}
\newcommand{\ee}{\end{equation}}
\newcommand{\ba}{\begin{eqnarray}}
\newcommand{\ea}{\end{eqnarray}}

\newcommand{\orcid}[1]{\href{https://orcid.org/#1}{\includegraphics[scale=0.08]{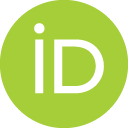}}}

\DeclareGraphicsExtensions{.pdf,.png}

\title[Biased Einstein radii]{Understanding the large inferred Einstein radii of observed low-mass galaxy clusters}

\author[A.\ Robertson \etal]{Andrew Robertson\thanks{e-mail: {\tt andrew.robertson@durham.ac.uk}}\orcid{0000-0002-0086-0524}, 
Richard Massey\orcid{0000-0002-6085-3780} and Vincent Eke\orcid{0000-0001-5416-8675}\\Institute for Computational Cosmology, Durham University, South Road, Durham DH1 3LE, UK\\
}

\begin{document}

\maketitle

\label{firstpage}

\begin{abstract}


We assess a claim that observed galaxy clusters with mass $\sim$$10^{14} \msun$ are more centrally concentrated than predicted in $\Lambda$CDM.
We generate mock strong gravitational lensing observations, taking the lenses from a cosmological hydrodynamical simulation, and analyse them in the same way as the real Universe. The observed and simulated lensing arcs are consistent with one another, with three main effects responsible for the previously claimed inconsistency. First, galaxy clusters containing baryonic matter have higher central densities than their counterparts simulated with only dark matter. 
Second, a sample of clusters selected because of the presence of pronounced gravitational lensing arcs preferentially finds centrally concentrated clusters with large Einstein radii. 
Third, lensed arcs are usually straighter than critical curves, and the chosen image analysis method (fitting circles through the arcs) overestimates the Einstein radii. 
After accounting for these three effects, $\Lambda$CDM predicts that galaxy clusters should produce giant lensing arcs that match those in the observed Universe.

\end{abstract}


\section{Introduction}

Combined strong and weak gravitational lensing analyses of galaxy clusters have demonstrated that NFW density profiles \citep[which are the predicted density profiles for dark matter haloes in a $\Lambda$CDM universe,][]{1997ApJ...490..493N} can explain the observed lensing, but that the inferred concentration parameters are often higher than those found in $N$-body simulations \citep[e.g.][]{2008ApJ...685L...9B,2010MNRAS.403.2077S, 2012MNRAS.420.3213O}. This is especially true for low mass clusters, with observed samples of more massive clusters having inferred concentrations in line with $N$-body predictions \citep{2015ApJ...806....4M}. A related phenomenon is that the Einstein radii, $\theta_\mathrm{E}$, of observed low-mass clusters are larger than predicted in $\Lambda$CDM \citep[][hereafter \citetalias{2012ApJ...761....1W}]{2012ApJ...761....1W}.

Expressed in terms of the concentration-mass relation, this over-concentration of low mass clusters results in a steeper observed dependence of halo concentration with mass than is predicted from $N$-body simulations \citep{2012MNRAS.420.3213O}. \citet{2013MNRAS.436..503A} have pointed out that observationally inferred $M_{200} - c$ relations typically follow the slope in the covariance between $M_{200}$ and $c$ that comes from a strong-lensing measurement. Strong lensing provides an estimate of the projected mass within the Einstein radius. For a given \emph{Einstein mass}, the concentration must be higher if the total halo mass is lower. As such, systems in which $M_{200}$ is overestimated will have low inferred concentrations, and systems in which $M_{200}$ is underestimated will have high inferred concentrations. Performing a Bayesian hierarchical inference in which they fit for the concentration--mass relation as well as for their underlying distribution of halo masses, \citet{2013MNRAS.436..503A} find that observations are in fact consistent with the concentration--mass relations found in simulations.

A direct comparison with observed quantities, i.e.\ predicting the observables from the simulations rather than inferring physical quantities (such as $M_{200}$ and $c$) from the observations, can circumvent some of the problems identified by \citet{2013MNRAS.436..503A}. \citetalias{2012ApJ...761....1W} found the Einstein radii of $M_{200} = 10^{14} - 10^{15} \msun$ clusters larger than expected for NFW profiles with the same masses as their clusters (measured with cluster richness), especially for the least massive haloes. Their requirement that systems need obvious strong lensing arcs biases them towards a sample of efficient lenses, which at fixed mass means preferentially finding the most concentrated systems, with the largest Einstein radii. However, even taking into account the lensing selection \citep[using a model from][]{2012MNRAS.420.3213O}, their Einstein radii are larger then expected \citepalias{2012ApJ...761....1W}.

In this paper we address the mismatch between observed and predicted Einstein radii, using mock strong lensing observations made from hydrodynamical simulations. We find that there are three effects that cause a discrepancy between the observed Einstein radii and the NFW predictions, and that taking all these effects into account the simulations and observations are in good agreement. These effects are that baryonic physics leads to increased central densities and so increased Einstein radii, that a lensing-selected sample preferentially contains the most efficient gravitational lenses (as already noted by \citetalias{2012ApJ...761....1W}), and that the method employed by \citetalias{2012ApJ...761....1W} to measure Einstein radii produces results that are biased towards large values. 

This paper is organised as follows. In Section~\ref{sect:sims} we discuss the hydrodynamical simulations and the methods we employ to generate mock lensing data from them. We then present the results of comparing these mock observations with the real observations in Section~\ref{sect:results}. We discuss the sensitivity of our results to the numerical and physical parameters that we adopted during our analysis in Section~\ref{sect:convergence}, before concluding in Section~\ref{sect:conclusions}.


\section{Gravitational lensing from simulated galaxy clusters}
\label{sect:sims}

For an axisymmetric lens with a suitably high central density, a small source directly behind the centre of the lens will be gravitationally lensed and will appear as a ring centred on the lens centre. Starting from the deflection angle for light passing a point mass, it can be shown that the radius of this ring is the radius at which the mean enclosed projected density is equal to the critical surface density for lensing, $\Sigma_\mathrm{crit}$. This is defined as
\begin{equation}
\Sigma_\mathrm{crit} = \frac{c^2}{4 \pi G} \frac{D_\mathrm{s}}{D_\mathrm{l}D_\mathrm{ls}}.
\label{sigma_crit}
\end{equation}
Here, $D_\mathrm{s}$, $D_\mathrm{l}$, and $D_\mathrm{ls}$ are the angular diameter distances between the observer and source, observer and lens, and lens and source respectively.

\subsection{Lensing by NFW haloes}

An NFW halo \citep{1997ApJ...490..493N} has a 3D density profile 
\begin{equation}
\frac{\rho(r)}{\rho_\mathrm{crit}} = \frac{\delta_\mathrm{NFW}}{(r/r_\mathrm{s})(1+r/r_\mathrm{s})^2},
\end{equation}
where $r_\mathrm{s}$ is the scale radius, $\rho_\mathrm{crit}=3H^2/8\pi G$ is the critical density of the universe, and $\delta_\mathrm{NFW}$ is a dimensionless characteristic density that can be related to the halo \emph{concentration}, $c$, through $\delta_\mathrm{NFW} = \frac{200}{3} \, c^3 / \left[ \ln (1+c)-c/(1+c) \right]$. Calculating the Einstein radius of such a halo requires that we integrate the 3D density along lines of sight at different impact parameters to get the projected surface density profile, $\Sigma(R)$, where $R$ is a 2D distance from the halo centre. From this we can find the radius within which the mean enclosed surface density, $\bar{\Sigma}(R)$, is equal to $\Sigma_\mathrm{crit}$, which is then the Einstein radius, $R_\mathrm{E}$. An analytical equation exists for $\bar{\Sigma}(R)$ of an NFW profile, but it takes a complicated form. We point the reader to equation 13 of \citet{2000ApJ...534...34W} if they wish to see it. We use the analytical form for $\bar{\Sigma}(R)$ to find where it is equal to $\Sigma_\mathrm{crit}$, and so to find the Einstein radius.

In the left hand panel of Fig.~\ref{fig:einstein_radii} we plot the Einstein radius as a function of halo mass, assuming that the NFW concentrations follow the \citet{2016MNRAS.460.1214L} concentration-mass relation at the lens redshift. Note that the physical Einstein radius, $R_\mathrm{E}$, has been converted into an angular Einstein radius, $\theta_\mathrm{E}$, by dividing by the angular diameter distance to the lens redshift, which is assumed to be $z_\mathrm{l} = 0.375$ throughout this paper.

\subsection{The BAHAMAS simulations}

To go beyond the NFW prediction, we use a hydrodynamical simulation from the \bahamas project \citep{2017MNRAS.465.2936M,McCarthy2018}. \bahamas was run using a modified version of the {\sc Gadget-3} code \citep{Springel2005}.  The simulations include subgrid treatments for metal-dependent radiative cooling \citep{Wiersma2009a}, star formation \citep{Schaye2008}, stellar evolution and chemodynamics \citep{Wiersma2009b}, and stellar and AGN feedback \citep{DallaVecchia2008,Booth2009}, developed as part of the OWLS project (see \citealt{Schaye2010} and references therein). The simulation we use is of a periodic box, $400 \, h^{-1} \, \mpc$ on a side, with $2 \times 1024^3$ particles. The simulation employs a \textit{WMAP} 9-yr cosmology\footnote{With $\Omega_\mathrm{m}=0.2793$, $\Omega_\mathrm{b}=0.0463$, $\Omega_\mathrm{\Lambda}=0.7207$, $\sigma_8 = 0.812$, $n_\mathrm{s} = 0.972$ and $h = 0.700$.} \citep{2013ApJS..208...19H}, and has dark matter (DM) and baryon particle masses of $\num{5.5e9} \msun$ and $\num{1.1e9} \msun$, respectively. The Plummer-equivalent gravitational softening length is $5.7 \kpc$ in physical coordinates below $z=3$ and is fixed in comoving coordinates at higher redshifts. We include all haloes with $M_{200} > 10^{14} \msun$ in our analysis, which leads to a sample of $1,040$ haloes at our lens redshift of $0.375$.

\subsection{Calculating deflection angles}
\label{sect:deflection_angles}

The key quantity required to do mock gravitational lensing with our simulated clusters is a deflection angle field. This describes the deflection of light rays as they pass through the simulated system, and so provides a mapping from the observed (\emph{lens plane}) coordinates, back to locations in the \emph{source plane}. The method used to generate deflection angle fields from our simulated clusters is the same as in \citet[][hereafter \citetalias{2019MNRAS.488.3646R}]{2019MNRAS.488.3646R}. In this work we adjust the values of some numerical parameters, particularly those related to the resolution of the 2D density field from which the deflection angles were calculated. The reason for this change is that a higher resolution map of deflection angles is required to produce realistic lensed arcs (as we do in this work) than is required simply to map out the tangential critical curve (as was done in \citetalias{2019MNRAS.488.3646R}). In this section we summarise the method, and state the key numerical parameters. For full details about the method see \citetalias{2019MNRAS.488.3646R}.

Our method begins by generating a projected density map of each simulated cluster, projecting the cluster along the simulation $z$-axis. We use an adaptive triangular shaped cloud scheme (ATSC), where each particle's mass is smoothed out in both the $x$ and $y$ directions by a triangular kernel with a full width of $2 \, r_{32}$, where $r_{32}$ is the 3D distance to a particle's 32nd nearest neighbour of the same particle species (so dark matter, gas and stars are each treated separately).

Dividing the projected surface density, $\Sigma$, by $\Sigma_\mathrm{crit}$, we get the dimensionless convergence field, $\kappa$. Note that for all lensing calculations in this paper we use the same  \textit{WMAP} 9-yr cosmology \citep{2013ApJS..208...19H} as used to run the simulations, with an assumed source redshift of $z_\mathrm{s} = 2$, and lens redshift $z_\mathrm{l} = 0.375$. 
Both $\kappa$ and the deflection angle field, $\vect{\alpha}$, depend on spatial derivatives of the projected gravitational potential. This means that the relationship between the Fourier transforms of $\kappa$ and $\vect{\alpha}$ is a simple one, and we calculate $\vect{\alpha}$ from $\kappa$ using discrete Fourier transforms \citepalias{2019MNRAS.488.3646R}. 

The main change from \citetalias{2019MNRAS.488.3646R} is that our 2D density maps are higher resolution (1024 pixels on a side, but now covering $2\times2 \mpc^2$, down from $4\times 4 \mpc^2$). The only other change is that the smoothing scale of each particle in the ATSC scheme uses the distance to the 32nd (rather than 8th) nearest neighbour, because this reduces the noise in our deflection-angle maps.

\begin{figure*}
  \centering
\begin{subfigure}[b]{0.52\linewidth}
    \centering
\includegraphics[width=\textwidth]{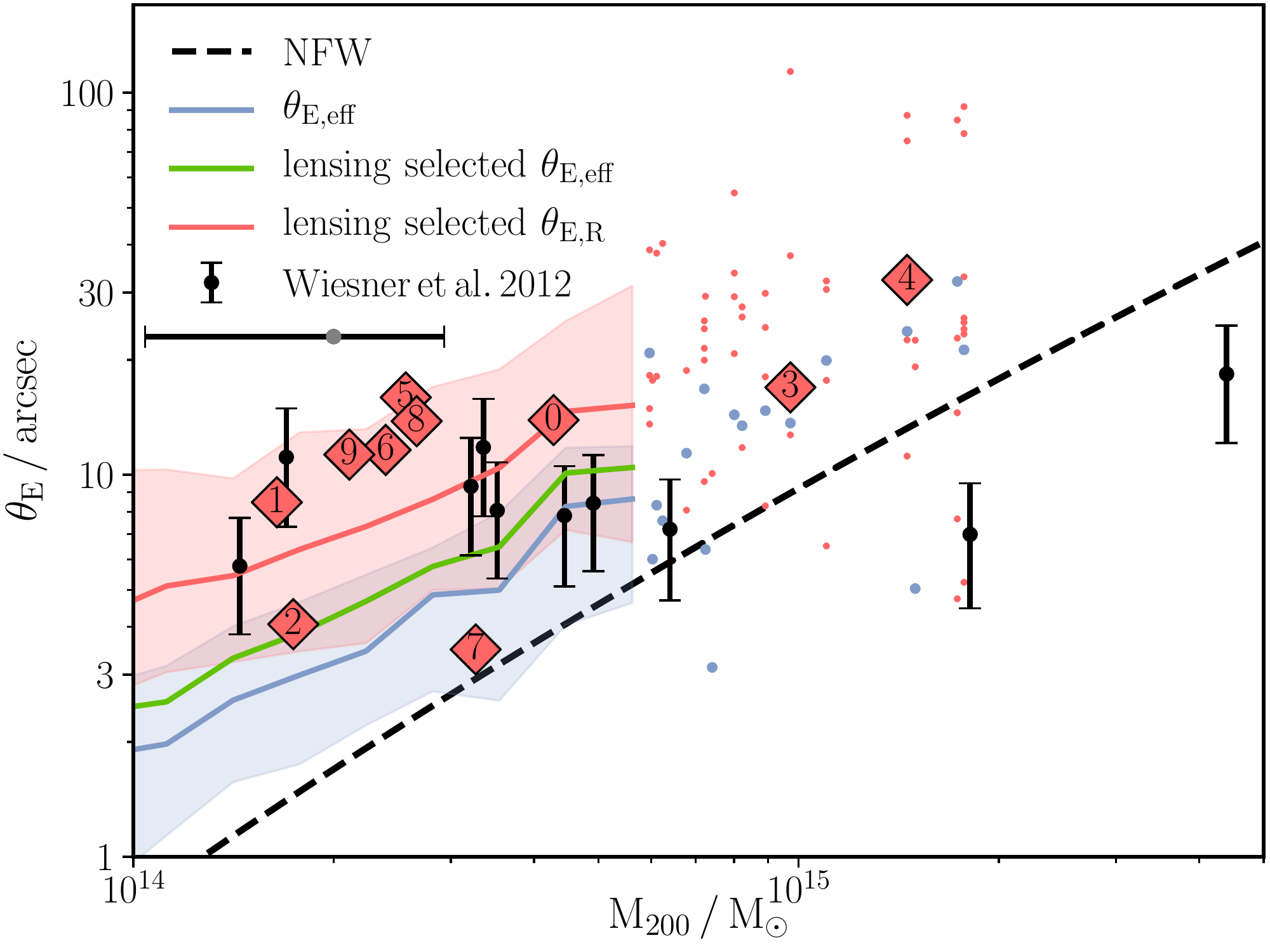}
\end{subfigure}%
\begin{subfigure}[b]{0.48\linewidth}
    \centering
    \includegraphics[clip, trim=0 0 0 0, width=0.95\textwidth]{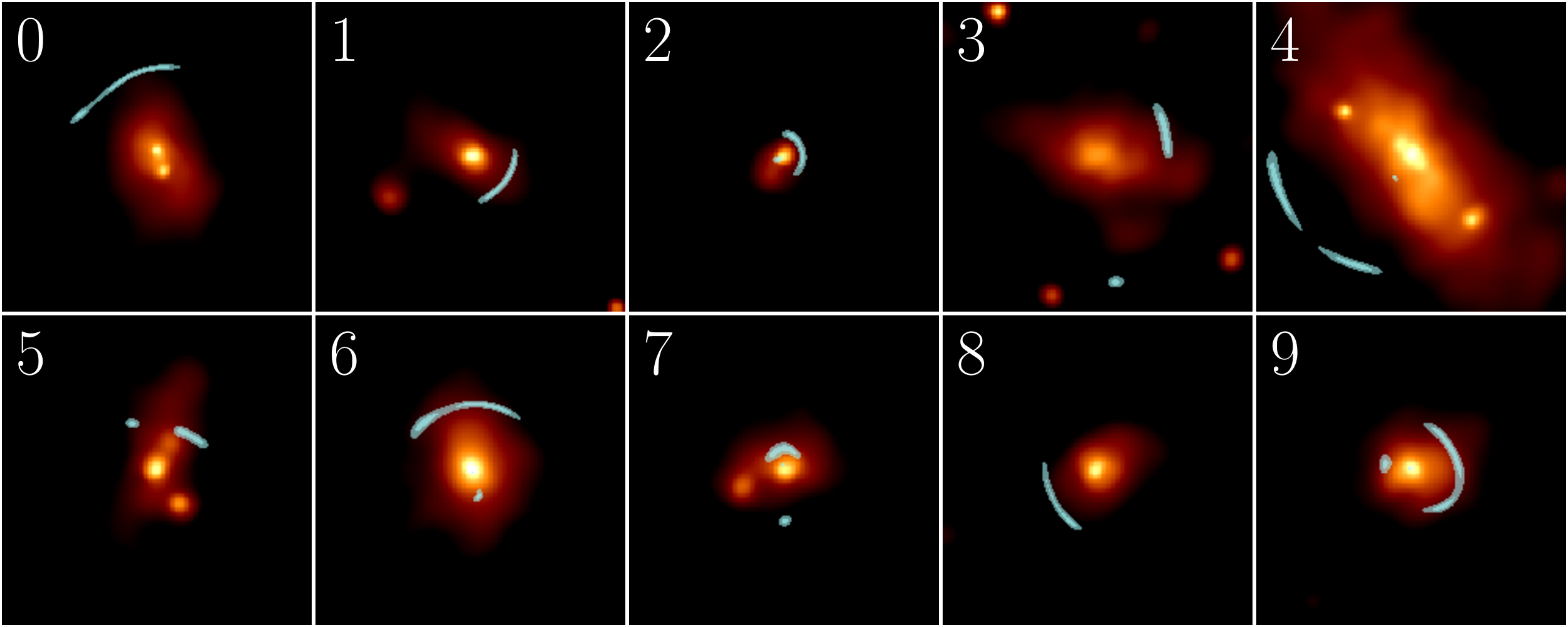}
\includegraphics[clip, trim=0 -50pt 0 0, width=0.95\textwidth]{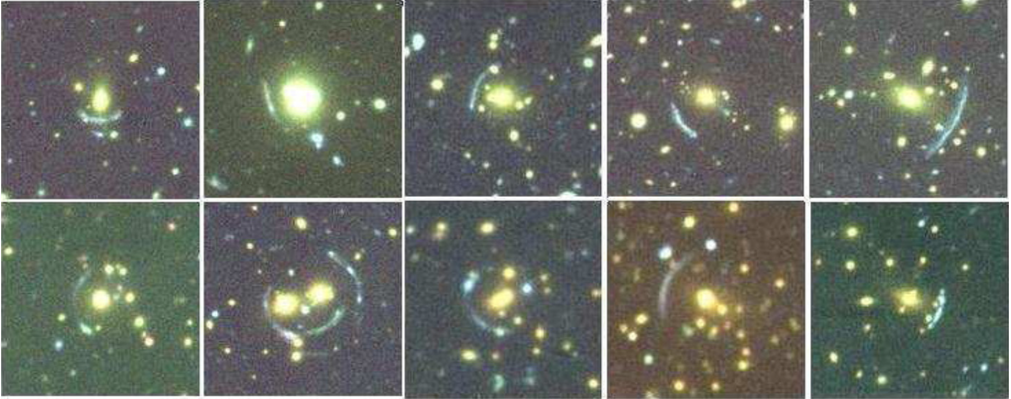}
  \end{subfigure} 
  \caption{Left: the Einstein radius as a function of halo mass for haloes at $z_\mathrm{l} = 0.375$ with a source redshift of $z_\mathrm{s} = 2$. The black dashed line shows the prediction for NFW haloes following the \citet{2016MNRAS.460.1214L} concentration-mass-redshift relation. The blue line shows the median $\effein$ from the \bahamas simulations, with the shaded region covering the 16th-84th percentiles. The green line shows the median $\effein$ from \bahamas when weighting each cluster by the number of highly magnified arcs that it produces. The red line shows the median Einstein radius estimated from the curvature of highly magnified arcs. As per the green line, each arc contributes equally to the median for the red line (as opposed to each cluster receiving equal weight), and the shaded region again covers the 16th-84th percentiles. The black points are measurements by \citetalias{2012ApJ...761....1W} based on the curvature of observed arcs, and so should be compared with the red line. The halo masses for the observed systems are measured from cluster richness and are quite uncertain, the typical uncertainty is plotted below the legend. At the high mass end we plot individual clusters (blue points) and arcs (red points), instead of red and blue lines and shaded regions. Top right: $49'' \times 49''$ images of simulated strong lensing arcs, with the arcs shown in blue on top of a stellar-mass map. The numbers in the top left of each map correspond to the numbers in red diamonds in the left panel. Bottom right: the observed strong lensing arcs from \citetalias{2012ApJ...761....1W} with $49'' \times 49''$ images produced from g, r and i filters (\textcopyright AAS, reproduced with permission).}
  \label{fig:einstein_radii}
\end{figure*}

\subsection{Effective Einstein radii}
\label{sect:EffEinRad}


While gravitational lensing preserves surface brightness, it can magnify background sources by increasing their area. Regions of the lens plane where the magnification is infinite are known as critical curves. The magnification is given by
\begin{equation}
\mu = \frac{1}{(1 - \kappa - \gamma)(1 - \kappa + \gamma)},
\label{eq:mu}
\end{equation}
where $\gamma$ is the magnitude of the gravitational shear. As the shear is also given by spatial derivatives of the projected gravitational potential \citep[e.g.][]{MeneghettiLensing}, it can be calculated from $\kappa$ in a similar manner to $\vect{\alpha}$ above.

Equation~\eqref{eq:mu} leads to two distinct types of critical curves: \emph{radial critical curves} appear where $1 - \kappa + \gamma = 0$, while \emph{tangential critical curves} occur where $1 - \kappa - \gamma = 0$, and lead to images stretched tangentially to the critical curve. For axisymmetric lenses, the latter of these is a circle with a radius that by definition is the Einstein radius, $\theta_\mathrm{E}$.

The definition of $\theta_\mathrm{E}$ can be extended to a general lens -- for which the tangential critical curve need not be circular -- by using the effective Einstein radius, $\effein$. This is the radius of a circle that encloses an area equal to the area enclosed by the tangential critical curve. In the left hand panel of Fig.~\ref{fig:einstein_radii} we plot the $\effein$ values for our simulated clusters as the blue line and points. These typically lie above the NFW prediction, reflecting the fact that the density in the centre of haloes is enhanced over the DM-only prediction due to both the baryonic mass itself, and the contracting effect it has on the DM distribution \citep{2004ApJ...616...16G}, and that triaxiality and/or a complex merging state can enhance the Einstein radii of galaxy clusters \citep[e.g.][]{2012A&A...547A..66R}.

\subsection{Highly magnified arcs and their curvature radii}
\label{sect:arc_curvature}

\citetalias{2012ApJ...761....1W} measured the Einstein radii of clusters from the properties of observed lensing arcs. In order to compare our simulations with the observation, we therefore need to generate lensing arcs from our simulated clusters. Our method to do this is similar to the method used in \citet{2001MNRAS.325..435M}. We distribute sources on a regular grid in the source plane, with $256^2$ sources behind each lens, covering an area of the source plane that is $3 \times 3$ arcmin$^2$. We consider one source at a time when generating and analysing mock lensed images, and use a much higher source-density than for any realistic population of sources to efficiently explore the possible lensing arcs produced by a given lensing mass distribution.

Each source is modelled as an ellipse, with axis ratio $q$, and an area equal to that of a circle with radius $r_\mathrm{source}$. For each source, we find all of the points on a regular grid in the lens plane that when mapped to the source plane are enclosed by the boundary of the source. For this purpose we use a higher resolution grid than was used for the calculation of the deflection angles, with a grid spacing of $0.02 \, \mathrm{arcsec}$. The deflection angles on this high-resolution grid are calculated using bilinear interpolation on the coarser deflection angle grid.

The lens-plane points that map to a location inside the source are split into sets that are contiguous in the lens plane (there can be distinct contiguous sets as some sources are multiply imaged), which are the individual lensing arcs. \citetalias{2012ApJ...761....1W} measured the Einstein radii of individual clusters by assuming that they are equal to the radius of curvature of bright lensing arcs. To compare with this, we need to determine the properties of each simulated arc, which we do by:
\begin{itemize}[nosep]
\item Finding the image point (a) that is closest to the source centre when mapped back to the source plane.
\item Finding the image point (b) that is farthest from (a).
\item Finding the image point (c) that is farthest from (b).
\item Fitting a circle through the three points; (a), (b) and (c), with the radius of this circle being the Einstein radius as measured from arc curvature, $\theta_\mathrm{E,R}$.
\item Determining the image area, $A$, from the number of image points and the lens-plane grid spacing.
\item Calculating the magnification of the image, $\mu$, from the ratio of $A$ to the area the source would cover in the absence of gravitational lensing.
\end{itemize}

Our sources are geometric objects into which points in the lens plane either map inside or outside. Compared with the lensing of real galaxies, which have a spatially varying surface brightness distribution, this may seem simplistic. However, gravitational lensing does not alter surface brightness, making these geometric sources a good approximation for the purpose of measuring the shapes of arcs. For an image with a particular surface brightness limit, $S$, the perimeter of a detected arc maps back to the isophote of the source galaxy with surface brightness $S$. So the perimeters of our simulated lensing arcs will look like the perimeters of observed lensing arcs, so long as the perimeters of our sources look like the (unlensed) isophotes of real lensed galaxies, which are usually well modelled as ellipses.

For our sources, $q$ was drawn from a uniform distribution between 0.4 and 1, in rough agreement with observed high-redshift galaxies \citep{2014ApJ...792L...6V}. We varied $r_\mathrm{source}$ in the range $1$ to $4 \kpc$, corresponding to $0.22$ to $0.87$ arcsec. For reference, the mean half-light radius of z=2 galaxies (in the rest-frame UV) measured by \citet{2004ApJ...600L.107F} is roughly 0.4 arcsec. Our results shown in Fig.~\ref{fig:einstein_radii} are for our fiducial source radius of $2 \kpc$.

\subsection{Identifying strongly lensed arcs}

In order to compare the results from our simulation with the observed sample, we need to include the selection effects that went into the \citetalias{2012ApJ...761....1W} sample. This observed sample of 10 strong-lensing galaxy clusters was compiled from visually inspecting images for likely lensed arcs. There were two samples of images that were inspected \citep{2009ApJ...696L..61K}, based on two different searches of the Sloan Digital Sky Survey Data Release Five \citep[SDSS DR5,][]{2007ApJS..172..634A}. The first search was for blue objects ($g - r < 1$ and $r - i < 1$) around catalogues of Luminous Red Galaxies and Brightest Cluster Galaxies, with the second a catalogue of suspected merging galaxies generated using the method described in \citet{2004AJ....127.1883A}.

This complicated selection function, including human inspection, is difficult to reproduce. As a proxy for the selection of visually identifiable lensing arcs, we impose a minimum magnification that a lensed arc must have to be included in our sample. We use a fiducial value of $\mu_\mathrm{min} = 16$, which we found produced samples of arcs that are similar in appearance to the observed arcs (see Fig.~\ref{fig:einstein_radii}). In Fig.~\ref{fig:convergence} we show that our results are only mildly affected by using a $\mu_\mathrm{min}$ of $8$ or $32$ instead.




\section{Results and Discussion}
\label{sect:results}

There are three primary reasons why \citetalias{2012ApJ...761....1W} measured Einstein radii that mainly lie above the prediction for NFW profiles that follow the median concentration-mass relation. The first of these is encapsulated in the blue line in Fig.~\ref{fig:einstein_radii}, which shows $\effein$ as a function of $M_{200}$ for haloes from \bahamas. At a given halo mass, the NFW prediction lies roughly along the 16th percentile line for $\effein$ from the simulated mass distributions, so the bulk of haloes are more efficient lenses than a spherically symmetric NFW profile with the median concentration predicted by DM-only simulations. As discussed in Section~\ref{sect:EffEinRad} this reflects the fact that departures from spherical symmetry generally enhance gravitational lensing, and that the cooling of gas into the centre of DM haloes increases the total density at the centre of haloes compared with the DM-only case.

The second reason for \citetalias{2012ApJ...761....1W} measuring large Einstein radii is that their sample is selected based on the presence of obvious strong lensing arcs. At fixed mass, more centrally concentrated mass distributions have larger Einstein radii and produce more strong lensing than their less concentrated counterparts \citep{2007A&A...473..715F, 2012MNRAS.420.3213O}. This means that haloes that appear in a lensing-selected sample will be biased towards large Einstein radii. To demonstrate this with our simulated haloes, we identified all source and lens combinations that produce an arc with $\mu > 16$ as being candidates that could have entered the \citetalias{2012ApJ...761....1W} sample. The green line in Fig.~\ref{fig:einstein_radii} is then the median $\effein$ as a function of $M_{200}$, where a halo's weight in the median is given by the number of sources it has that produce a $\mu > 16$ arc. What this means explicitly, is that within each $M_{200}$ bin, each halo's $\effein$ is included in a list of $\effein$ a number of times that is equal to the number of sources it lenses to produce a $\mu > 16$ arc. The median of the $\effein$ in this list then gives the lensing-selected $\effein$ value for this mass bin. Note that by using $\effein$ the green line also includes the effects that lead the blue line to differ from the NFW prediction.

\begin{figure}
        \centering
        \includegraphics[width=\columnwidth]{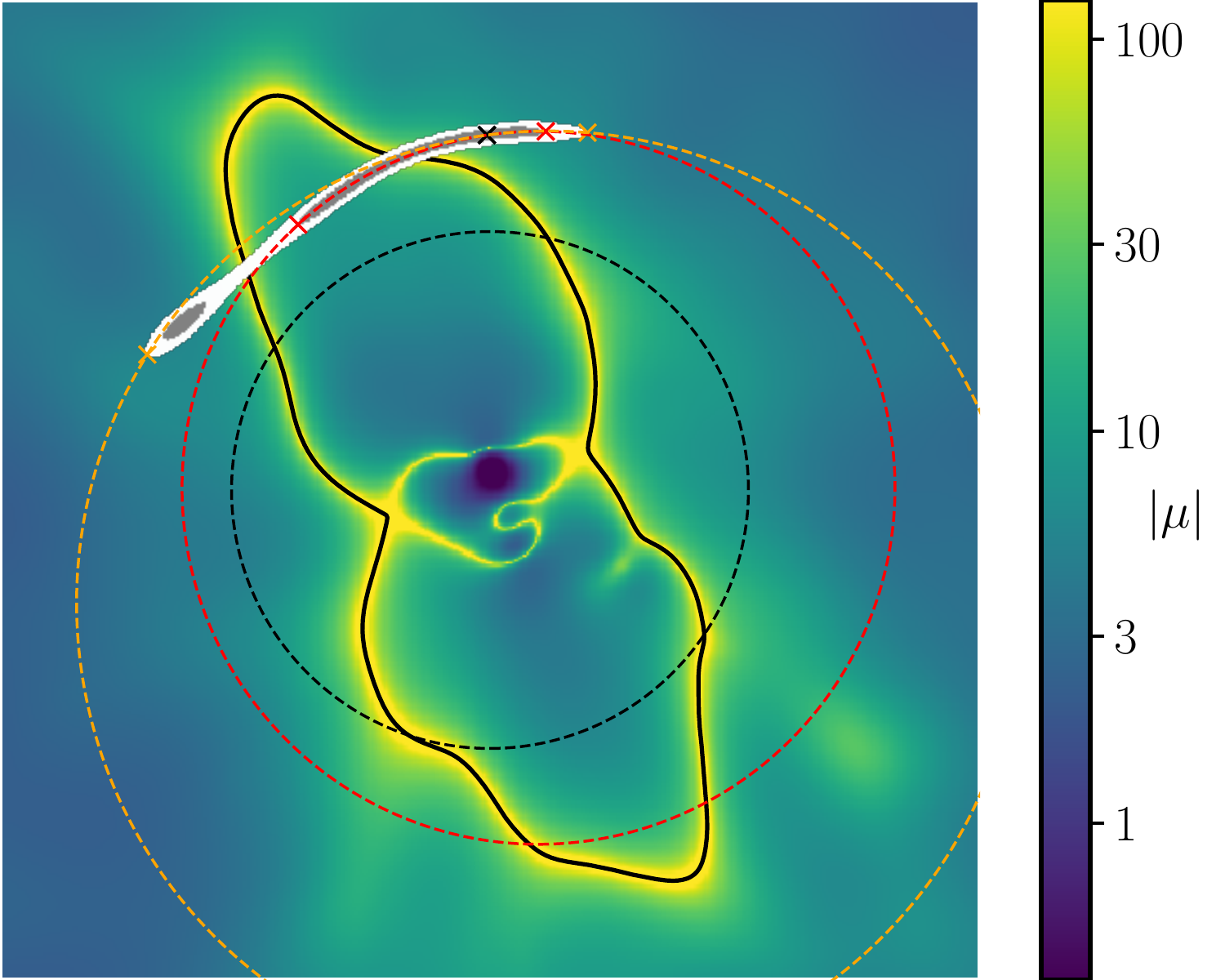}     
	\caption{A $40'' \times 40''$ magnification map of the system that produced `arc 0' in Fig.~\ref{fig:einstein_radii}, with the tangential critical curve drawn as a black solid line, and the effective Einstein radius, $\effein$, indicated by the black dashed circle. The same lensed source from Fig.~\ref{fig:einstein_radii} is overlaid, shown with a source radius of both $4 \kpc$ in white (where it is one long arc) and $2 \kpc$ in grey (where it splits into two arcs). The black cross is the location of the lens-plane pixel whose centre is closest to the source's centre when mapped into the source plane. The red crosses show the two extremities of the $r_\mathrm{source}=2 \kpc$ arc, while the orange crosses show the same for the $r_\mathrm{source}=4 \kpc$ arc. The red and orange dashed lines are the circles that pass through the black cross and their respective coloured crosses, whose radii define $\theta_\mathrm{E,R}$. For both source sizes $\theta_\mathrm{E,R}$ is larger than $\effein$ which is true of most highly magnified arcs.}
	\label{fig:example}
\end{figure}

The third reason why \citetalias{2012ApJ...761....1W} found larger values for $\theta_\mathrm{E}$ than is predicted by naive models, is the way in which they measured the Einstein radii from their observed arcs. Specifically, they fitted circles to the visible arcs and assumed that the radii of those circles were the Einstein radii of the lenses. We carried out this same procedure on the simulated arcs that met our selection criteria, using the method described in Section~\ref{sect:arc_curvature}. An example of this procedure can be seen in Fig.~\ref{fig:example}, where we show a strong lensing arc overlaid onto a magnification map of one of our simulated galaxy clusters. This example is typical of highly magnified arcs, with a radius of curvature that is larger than $\effein$. The red line in Fig.~\ref{fig:einstein_radii} includes this effect as well as the preceding two, and is now in reasonable agreement with the observed systems.

To give a visual indication that our simulated arcs are similar to those from the observations, on the right hand side of Fig.~\ref{fig:einstein_radii} we show colour images of the 10 observed systems in \citetalias{2012ApJ...761....1W} as well as mock images made from 10 random arcs that met our $\mu > 16$ criterion. In terms of the stellar mass distribution within the lens, \bahamas is lacking many of the smaller galaxies that can be seen in the observed systems. This is unsurprising, given the resolution of \bahamas means that galaxies with stellar masses below $ 10^{10} \msun$ are resolved with fewer than 20 star particles. However, the lensed arcs appear visually similar to those in the observations, which suggests that our choice of source size, as well as our magnification threshold required to `detect' our lensed sources are reasonable choices for making a comparison with the observations. It is hard to make definitive statements, given the small number of systems in the \citetalias{2012ApJ...761....1W} sample, but certainly there does not appear to be evidence of an `overconcentration problem' when comparing the observations with their counterparts generated from the \bahamas simulations.

The effects that lead to differences between the NFW predictions and the $\theta_\mathrm{E,eff}$ values of our simulated clusters in Fig.~\ref{fig:einstein_radii} can be partially captured by using a concentration-mass relation fit to the total density profiles from the \bahamas simulations. However, the density profiles of the clusters in the \bahamas simulations systematically differ from NFW profiles, such that the best-fit NFW concentrations are sensitive to how the fitting is done. Also, for a reasonable choice of fitting procedure,\footnote{Specifically, we fit NFW profiles by minimising the sum of $\left( \log_{10} \rho_\mathrm{sim} (r_i) - \log_{10} \rho_\mathrm{NFW} (r_i) \right)^2$ with 42  logarithmically-spaced $r_i$ between $0.01 \, r_{200}$ and $r_{200}$.} using a concentration-mass relation fit to our hydodynamical simulations explains less than half of the difference between the NFW and $\theta_\mathrm{E,eff}$ lines in Fig.~\ref{fig:einstein_radii}. We therefore do not pursue NFW profiles with modified concentrations as a way of understanding the $\theta_\mathrm{E,eff}$ values of simulated haloes.

We note that while our paper indicates that the observed systems in \citetalias{2012ApJ...761....1W} are consistent with the \bahamas hydrodynamical $\Lambda$CDM simulation, there is recent and ongoing work to produce larger samples of bright lensing arcs that can be used to study the mass profiles of galaxy groups and clusters \citep[e.g.][]{2017ApJS..232...15D, 2020ApJS..247...12S}. These larger samples will provide improved statistics, and can be analysed using more sophisticated lens-modelling methods than measuring arc curvature, in order to derive constraints on cosmology. This will also require more detailed theoretical study, for example to understand the degeneracies between cosmological parameters and different implementations of baryonic physics within simulations.


\section{Sensitivity to adopted parameters}
\label{sect:convergence}

In order to trust the comparison between our mock lensing arcs and the observed systems from \citetalias{2012ApJ...761....1W}, we need to verify that they are not sensitive to the numerical parameters of the hydrodynamical simulation we used, or to the choices we made in our mock lensing procedure. We investigate these in this section, first mimicking the effects of having a lower resolution simulation, and then seeing how our results depend on the source size, selection of arcs, and lens and source redshifts.

\subsection{Effects of simulation resolution}



\begin{figure}
        \centering
        \includegraphics[width=\columnwidth]{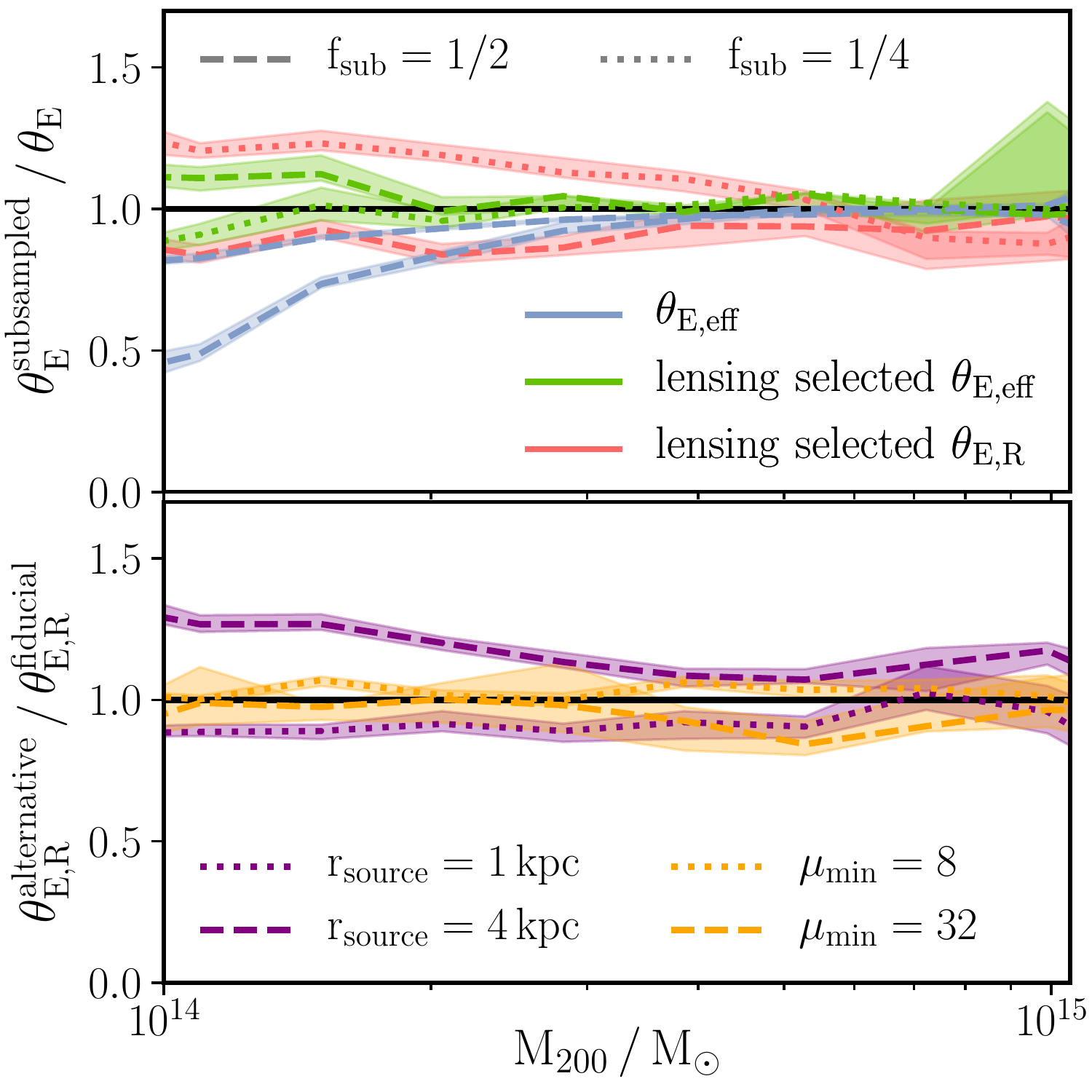}     
	\caption{The dependence of our results on the simulation resolution (top panel) and on our choice of source size and criterion for lensed arcs to be included in our sample (bottom panel). In all cases we plot the median Einstein radius as a function of halo mass, divided by its value with our fiducial setup. The shaded regions represent the 16th-84th percentile estimates of these ratios, from bootstrap resampling of our haloes. In the top panel we show how $\effein$, lensing-selected $\effein$, and $\theta_\mathrm{E,R}$ are affected as we decrease the particle sampling from our simulations by a factor of two or four. In the bottom panel we show how $\theta_\mathrm{E,R}$ is affected by an increase or decrease in the size of our sources, or by changes to the minimum magnification required for an arc to be included in our sample.}
	\label{fig:convergence}
\end{figure}

Determining how our results depend on the resolution of our simulations would ideally be done by redoing the analysis with simulations with different resolution. We do not have higher or lower resolution simulations with which we can compare, but we can mimic the effects of simulations with different resolutions by subsampling particles from our simulations. We generated lensing maps of all our haloes, when only using a fraction, $f_\mathrm{sub}$, of the simulation particles. When making these \emph{subsampled} maps, the mass of each particle was increased by $1/f_\mathrm{sub}$ to create a lower resolution version of the same simulated mass distribution. Using these subsampled simulations, we then carried out the same procedures as we had done to generate the blue, green and red lines in Fig.~\ref{fig:einstein_radii}, using our fiducial source radius of $2 \kpc$ and fiducial selection criterion of $\mu > 16$. The results of this process, relative to the results with the full simulation data are plotted in the top panel of Fig.~\ref{fig:convergence}.

In Fig.~\ref{fig:convergence} we can see that as we decrease $f_\mathrm{sub}$ (corresponding to lowering the resolution of our simulations) the median Einstein radius of low-$M_{200}$ haloes decreases. This is to be expected, because the method we employ to generate a projected density map (Section~\ref{sect:deflection_angles}) smooths the mass distribution on a scale that depends on the distance between particles and their neighbours. Lowering the resolution of the simulation increases this smoothing scale. Haloes whose Einstein radii are significantly larger than this smoothing scale will not be affected, but lower mass systems -- which typically have smaller Einstein radii -- will have their Einstein radii shrink as their mass distributions are smoothed on larger scales.

The reduction in $\effein$ as the simulation resolution is decreased is not reflected in the medians of either the lensing selected sample, or in the arc curvature radii, $\theta_\mathrm{E,R}$. This is because larger $\effein$ systems are the ones that are less affected by resolution. At fixed halo mass, the distribution of $\effein$ is approximately log-normal with a standard deviation of 0.24 dex.\footnote{This can be seen in the blue shaded region in Fig.~\ref{fig:einstein_radii} where the 16th-84th percentile range covers a factor of roughly three.} The probability to produce a highly magnified arc scales approximately as $\effein^{2.4}$ \citep{2013SSRv..177...31M}, which combined with 0.24 dex scatter would mean that systems in the top 16\% of the $\effein$ distribution produce over 60\% of the strong lensing arcs. These systems are the better resolved ones, and so when selecting systems based on their ability to produce highly magnified arcs, simulation resolution is less important than for a mass-selected sample of haloes. This suggests that our result that when taking into account selection effects and the method for measuring Einstein radii, the \citetalias{2012ApJ...761....1W} observations are consistent with $\Lambda$CDM simulations, is robust to changes in simulation resolution.

\subsection{Effects of source size}

Our adopted source radius of $2 \kpc$ for a $z=2$ source was selected as this is approximately the half-light radius of $z=2$ galaxies that are selected for being bright in the rest-frame UV \citep{2004ApJ...600L.107F}. However, the surface brightness to which \citetalias{2012ApJ...761....1W} can identify arcs may not correspond to the typical surface brightnesses at the half-light radii of the source galaxies, and a selection based on being lensed into a highly magnified arc may differ from that which created the \citet{2004ApJ...600L.107F} sample. To address this, we investigated how our results change with assumed source size.

In the bottom panel of Fig.~\ref{fig:convergence} we plot the median $\theta_\mathrm{E,R}$ for source radii of $1$ and $4 \kpc$, relative to the fiducial case. We find that the general trend is the same as in the specific case shown in Fig.~\ref{fig:example}, with $\theta_\mathrm{E,R}$ increasing with increasing source radius. For the observational sample, not all sources will have the same effective radius, but because factor of two changes in source size lead to only $10-20 \%$ changes in $\theta_\mathrm{E,R}$, our results are not very sensitive to our choice of fiducial source size. The fact that the simulated and observed arcs that we show on the right hand side of Fig.~\ref{fig:einstein_radii} look similar, suggests that our fiducial source size is a reasonable approximation to the source sizes in the observed sample.

\subsection{Effects of sample selection}

Another choice we made in our analysis that could affect the properties of our simulated arcs is the selection criterion for including arcs, for which we used $\mu > 16$. In the bottom panel of Fig.~\ref{fig:convergence} we show how our results change if we include all arcs with $\mu > 8$ or only those with $\mu > 32$. There is an indication that the typical curvature radii of our arcs decrease slightly with increasing arc magnification, but the size of this change is very small, so our results are insensitive to precisely how we select arcs to include in our sample.


\subsection{Effects of lens and source redshift}

Throughout this work we have used lens and source redshifts of 0.375 and 2 respectively. For the observed lenses, the lens redshifts varied from 0.26 to 0.56 and the source redshifts from 0.66 to 2.94, with our source and lens redshifts selected to be near the middle of these ranges.\footnote{The specific choice for the lens redshift was dictated by requiring that it coincided with a snapshot from \bahamas.} To see how we would expect our results to differ with a different choice of lensing geometry, we can consider the case of lensing by isothermal spheres. Although our lenses are not exactly isothermal spheres, they can be used to assess approximately how efficient different lensing geometries are. For an isothermal sphere, the Einstein radius is proportional to $D_\mathrm{s} / D_\mathrm{ls}$ \citep{1996astro.ph..6001N}. For our lensing geometry and assumed cosmology this ratio is 0.72, while for the lenses in \citetalias{2012ApJ...761....1W} the median value of this ratio is 0.65, with a standard deviation of 0.12. Although our lensing geometry is more efficient for lensing than the average geometry in the observations, the expected shift to the Einstein radii had we used this average geometry is only around 10\%.


\section{Conclusions}
\label{sect:conclusions}

We have shown that observations of ten giant lensing arcs as part of the Sloan Bright Arcs Survey \citep[][(\citetalias{2012ApJ...761....1W})]{2012ApJ...761....1W} are in agreement with predictions from $\Lambda$CDM. This is in contrast to a claim by \citetalias{2012ApJ...761....1W} that their lower mass clusters had larger Einstein radii than $\Lambda$CDM would predict. This statement is true when comparing the measured Einstein radii with those predicted for typical NFW profiles found from dark matter-only $\Lambda$CDM simulations, but there are three effects that explain why the inferred Einstein radii in \citetalias{2012ApJ...761....1W} lie above those predicted from NFW profiles following the median concentration-mass relation. These effects are that:
\begin{enumerate}[1.]
\item The total densities in the central regions of low-mass galaxy clusters are higher in hydrodynamical simulations than their counterparts in DM-only simulations, owing to both the mass in stars, and an increased DM density due to adiabatic contraction.
\item The observed systems were selected because they produced giant lensing arcs, which preferentially selects for systems with larger Einstein radii.
\item The method employed by \citetalias{2012ApJ...761....1W} to measure the Einstein radii (fitting circles through the lensed arcs), leads to larger Einstein radii than the true values. 
\end{enumerate}
In the left panel of Fig.~\ref{fig:einstein_radii}, the blue line captures effect 1, the green line the combination of effects 1 and 2, and the red line the combination of all three. Taking into account these three effects we find that the observed lensing arcs are in good agreement with what is predicted by \bahamas, a hydrodynamical simulation of a $\Lambda$CDM universe.

\section*{Acknowledgments}

Thanks to Ian McCarthy for allowing us to use the \bahamas simulations, and to Matthew Wiesner for allowing us to reproduce his Sloan Bright Arcs Survey images in the bottom-right of our Fig.~\ref{fig:einstein_radii}. AR is supported by the European Research Council's Horizon2020 project `EWC’ (award AMD-776247-6). RM acknowledges the support of a Royal Society University Research Fellowship. VRE acknowledges support from STFC grant ST/P000541/1. This work used the DiRAC Data Centric system at Durham University, operated by the Institute for Computational Cosmology on behalf of the STFC DiRAC HPC Facility (www.dirac.ac.uk). This equipment was funded by BIS National E-infrastructure capital grant ST/K00042X/1, STFC capital grants ST/H008519/1 and ST/K00087X/1, STFC DiRAC Operations grant  ST/K003267/1 and Durham University. DiRAC is part of the National E-Infrastructure.

\bibliographystyle{mnras}

\bibliography{bibliography}

\begin{thebibliography}{}
\makeatletter
\relax
\def\mn@urlcharsother{\let\do\@makeother \do\$\do\&\do\#\do\^\do\_\do\%\do\~}
\def\mn@doi{\begingroup\mn@urlcharsother \@ifnextchar [ {\mn@doi@}
  {\mn@doi@[]}}
\def\mn@doi@[#1]#2{\def\@tempa{#1}\ifx\@tempa\@empty \href
  {http://dx.doi.org/#2} {doi:#2}\else \href {http://dx.doi.org/#2} {#1}\fi
  \endgroup}
\def\mn@eprint#1#2{\mn@eprint@#1:#2::\@nil}
\def\mn@eprint@arXiv#1{\href {http://arxiv.org/abs/#1} {{\tt arXiv:#1}}}
\def\mn@eprint@dblp#1{\href {http://dblp.uni-trier.de/rec/bibtex/#1.xml}
  {dblp:#1}}
\def\mn@eprint@#1:#2:#3:#4\@nil{\def\@tempa {#1}\def\@tempb {#2}\def\@tempc
  {#3}\ifx \@tempc \@empty \let \@tempc \@tempb \let \@tempb \@tempa \fi \ifx
  \@tempb \@empty \def\@tempb {arXiv}\fi \@ifundefined
  {mn@eprint@\@tempb}{\@tempb:\@tempc}{\expandafter \expandafter \csname
  mn@eprint@\@tempb\endcsname \expandafter{\@tempc}}}

\bibitem[\protect\citeauthoryear{{Adelman-McCarthy} et~al.,}{{Adelman-McCarthy}
  et~al.}{2007}]{2007ApJS..172..634A}
{Adelman-McCarthy} J.~K.,  et~al., 2007, \mn@doi [\apjs] {10.1086/518864},
  \href {https://ui.adsabs.harvard.edu/abs/2007ApJS..172..634A} {172, 634}

\bibitem[\protect\citeauthoryear{{Allam}, {Tucker}, {Smith}, {Lee}, {Annis},
  {Lin}, {Karachentsev}  \& {Laubscher}}{{Allam}
  et~al.}{2004}]{2004AJ....127.1883A}
{Allam} S.~S.,  {Tucker} D.~L.,  {Smith} J.~A.,  {Lee} B.~C.,  {Annis} J.,
  {Lin} H.,  {Karachentsev} I.~D.,   {Laubscher} B.~E.,  2004, \mn@doi [\aj]
  {10.1086/381954}, \href
  {https://ui.adsabs.harvard.edu/abs/2004AJ....127.1883A} {127, 1883}

\bibitem[\protect\citeauthoryear{{Auger}, {Budzynski}, {Belokurov}, {Koposov}
  \& {McCarthy}}{{Auger} et~al.}{2013}]{2013MNRAS.436..503A}
{Auger} M.~W.,  {Budzynski} J.~M.,  {Belokurov} V.,  {Koposov} S.~E.,
  {McCarthy} I.~G.,  2013, \mn@doi [\mnras] {10.1093/mnras/stt1585}, \href
  {https://ui.adsabs.harvard.edu/abs/2013MNRAS.436..503A} {436, 503}

\bibitem[\protect\citeauthoryear{{Booth} \& {Schaye}}{{Booth} \&
  {Schaye}}{2009}]{Booth2009}
{Booth} C.~M.,  {Schaye} J.,  2009, \mn@doi [\mnras]
  {10.1111/j.1365-2966.2009.15043.x}, \href
  {http://adsabs.harvard.edu/abs/2009MNRAS.398...53B} {398, 53}

\bibitem[\protect\citeauthoryear{{Broadhurst}, {Umetsu}, {Medezinski}, {Oguri}
  \& {Rephaeli}}{{Broadhurst} et~al.}{2008}]{2008ApJ...685L...9B}
{Broadhurst} T.,  {Umetsu} K.,  {Medezinski} E.,  {Oguri} M.,   {Rephaeli} Y.,
  2008, \mn@doi [\apjl] {10.1086/592400}, \href
  {http://adsabs.harvard.edu/abs/2008ApJ...685L...9B} {685, L9}

\bibitem[\protect\citeauthoryear{{Dalla Vecchia} \& {Schaye}}{{Dalla Vecchia}
  \& {Schaye}}{2008}]{DallaVecchia2008}
{Dalla Vecchia} C.,  {Schaye} J.,  2008, \mn@doi [\mnras]
  {10.1111/j.1365-2966.2008.13322.x}, \href
  {http://adsabs.harvard.edu/abs/2008MNRAS.387.1431D} {387, 1431}

\bibitem[\protect\citeauthoryear{{Diehl} et~al.,}{{Diehl}
  et~al.}{2017}]{2017ApJS..232...15D}
{Diehl} H.~T.,  et~al., 2017, \mn@doi [\apjs] {10.3847/1538-4365/aa8667}, \href
  {https://ui.adsabs.harvard.edu/abs/2017ApJS..232...15D} {232, 15}

\bibitem[\protect\citeauthoryear{{Fedeli}, {Bartelmann}, {Meneghetti}  \&
  {Moscardini}}{{Fedeli} et~al.}{2007}]{2007A&A...473..715F}
{Fedeli} C.,  {Bartelmann} M.,  {Meneghetti} M.,   {Moscardini} L.,  2007,
  \mn@doi [\aap] {10.1051/0004-6361:20077926}, \href
  {https://ui.adsabs.harvard.edu/abs/2007A&A...473..715F} {473, 715}

\bibitem[\protect\citeauthoryear{{Ferguson} et~al.,}{{Ferguson}
  et~al.}{2004}]{2004ApJ...600L.107F}
{Ferguson} H.~C.,  et~al., 2004, \mn@doi [\apjl] {10.1086/378578}, \href
  {https://ui.adsabs.harvard.edu/abs/2004ApJ...600L.107F} {600, L107}

\bibitem[\protect\citeauthoryear{{Gnedin}, {Kravtsov}, {Klypin}  \&
  {Nagai}}{{Gnedin} et~al.}{2004}]{2004ApJ...616...16G}
{Gnedin} O.~Y.,  {Kravtsov} A.~V.,  {Klypin} A.~A.,   {Nagai} D.,  2004,
  \mn@doi [\apj] {10.1086/424914}, \href
  {https://ui.adsabs.harvard.edu/abs/2004ApJ...616...16G} {616, 16}

\bibitem[\protect\citeauthoryear{{Hinshaw} et~al.,}{{Hinshaw}
  et~al.}{2013}]{2013ApJS..208...19H}
{Hinshaw} G.,  et~al., 2013, \mn@doi [\apjs] {10.1088/0067-0049/208/2/19},
  \href {http://adsabs.harvard.edu/abs/2013ApJS..208...19H} {208, 19}

\bibitem[\protect\citeauthoryear{{Kubo}, {Allam}, {Annis}, {Buckley-Geer},
  {Diehl}, {Kubik}, {Lin}  \& {Tucker}}{{Kubo}
  et~al.}{2009}]{2009ApJ...696L..61K}
{Kubo} J.~M.,  {Allam} S.~S.,  {Annis} J.,  {Buckley-Geer} E.~J.,  {Diehl}
  H.~T.,  {Kubik} D.,  {Lin} H.,   {Tucker} D.,  2009, \mn@doi [\apjl]
  {10.1088/0004-637X/696/1/L61}, \href
  {https://ui.adsabs.harvard.edu/abs/2009ApJ...696L..61K} {696, L61}

\bibitem[\protect\citeauthoryear{{Ludlow}, {Bose}, {Angulo}, {Wang},
  {Hellwing}, {Navarro}, {Cole}  \& {Frenk}}{{Ludlow}
  et~al.}{2016}]{2016MNRAS.460.1214L}
{Ludlow} A.~D.,  {Bose} S.,  {Angulo} R.~E.,  {Wang} L.,  {Hellwing} W.~A.,
  {Navarro} J.~F.,  {Cole} S.,   {Frenk} C.~S.,  2016, \mn@doi [\mnras]
  {10.1093/mnras/stw1046}, \href
  {https://ui.adsabs.harvard.edu/abs/2016MNRAS.460.1214L} {460, 1214}

\bibitem[\protect\citeauthoryear{{McCarthy}, {Schaye}, {Bird}  \& {Le
  Brun}}{{McCarthy} et~al.}{2017}]{2017MNRAS.465.2936M}
{McCarthy} I.~G.,  {Schaye} J.,  {Bird} S.,   {Le Brun} A.~M.~C.,  2017,
  \mn@doi [\mnras] {10.1093/mnras/stw2792}, \href
  {http://adsabs.harvard.edu/abs/2017MNRAS.465.2936M} {465, 2936}

\bibitem[\protect\citeauthoryear{{McCarthy}, {Bird}, {Schaye},
  {Harnois-Deraps}, {Font}  \& {van Waerbeke}}{{McCarthy}
  et~al.}{2018}]{McCarthy2018}
{McCarthy} I.~G.,  {Bird} S.,  {Schaye} J.,  {Harnois-Deraps} J.,  {Font}
  A.~S.,   {van Waerbeke} L.,  2018, \mn@doi [\mnras] {10.1093/mnras/sty377},
  \href {http://cdsads.u-strasbg.fr/abs/2018MNRAS.476.2999M} {476, 2999}

\bibitem[\protect\citeauthoryear{Meneghetti}{Meneghetti}{2016}]{MeneghettiLensing}
Meneghetti M.,  2016, Introduction to Gravitational Lensing - Lecture scripts.
\url {http://www.ita.uni-heidelberg.de/~massimo/sub/Lectures/gl_all.pdf}

\bibitem[\protect\citeauthoryear{{Meneghetti}, {Yoshida}, {Bartelmann},
  {Moscardini}, {Springel}, {Tormen}  \& {White}}{{Meneghetti}
  et~al.}{2001}]{2001MNRAS.325..435M}
{Meneghetti} M.,  {Yoshida} N.,  {Bartelmann} M.,  {Moscardini} L.,  {Springel}
  V.,  {Tormen} G.,   {White} S.~D.~M.,  2001, \mn@doi [\mnras]
  {10.1046/j.1365-8711.2001.04477.x}, \href
  {http://adsabs.harvard.edu/abs/2001MNRAS.325..435M} {325, 435}

\bibitem[\protect\citeauthoryear{{Meneghetti}, {Bartelmann}, {Dahle}  \&
  {Limousin}}{{Meneghetti} et~al.}{2013}]{2013SSRv..177...31M}
{Meneghetti} M.,  {Bartelmann} M.,  {Dahle} H.,   {Limousin} M.,  2013, \mn@doi
  [\ssr] {10.1007/s11214-013-9981-x}, \href
  {https://ui.adsabs.harvard.edu/abs/2013SSRv..177...31M} {177, 31}

\bibitem[\protect\citeauthoryear{{Merten} et~al.,}{{Merten}
  et~al.}{2015}]{2015ApJ...806....4M}
{Merten} J.,  et~al., 2015, \mn@doi [\apj] {10.1088/0004-637X/806/1/4}, \href
  {http://adsabs.harvard.edu/abs/2015ApJ...806....4M} {806, 4}

\bibitem[\protect\citeauthoryear{{Narayan} \& {Bartelmann}}{{Narayan} \&
  {Bartelmann}}{1996}]{1996astro.ph..6001N}
{Narayan} R.,  {Bartelmann} M.,  1996, arXiv e-prints, \href
  {https://ui.adsabs.harvard.edu/abs/1996astro.ph..6001N} {pp
  astro--ph/9606001}

\bibitem[\protect\citeauthoryear{{Navarro}, {Frenk}  \& {White}}{{Navarro}
  et~al.}{1997}]{1997ApJ...490..493N}
{Navarro} J.~F.,  {Frenk} C.~S.,   {White} S.~D.~M.,  1997, \apj, \href
  {http://adsabs.harvard.edu/abs/1997ApJ...490..493N} {490, 493}

\bibitem[\protect\citeauthoryear{{Oguri}, {Bayliss}, {Dahle}, {Sharon},
  {Gladders}, {Natarajan}, {Hennawi}  \& {Koester}}{{Oguri}
  et~al.}{2012}]{2012MNRAS.420.3213O}
{Oguri} M.,  {Bayliss} M.~B.,  {Dahle} H.,  {Sharon} K.,  {Gladders} M.~D.,
  {Natarajan} P.,  {Hennawi} J.~F.,   {Koester} B.~P.,  2012, \mn@doi [\mnras]
  {10.1111/j.1365-2966.2011.20248.x}, \href
  {https://ui.adsabs.harvard.edu/abs/2012MNRAS.420.3213O} {420, 3213}

\bibitem[\protect\citeauthoryear{{Redlich}, {Bartelmann}, {Waizmann}  \&
  {Fedeli}}{{Redlich} et~al.}{2012}]{2012A&A...547A..66R}
{Redlich} M.,  {Bartelmann} M.,  {Waizmann} J.~C.,   {Fedeli} C.,  2012,
  \mn@doi [\aap] {10.1051/0004-6361/201219722}, \href
  {https://ui.adsabs.harvard.edu/abs/2012A&A...547A..66R} {547, A66}

\bibitem[\protect\citeauthoryear{{Robertson}, {Harvey}, {Massey}, {Eke},
  {McCarthy}, {Jauzac}, {Li}  \& {Schaye}}{{Robertson}
  et~al.}{2019}]{2019MNRAS.488.3646R}
{Robertson} A.,  {Harvey} D.,  {Massey} R.,  {Eke} V.,  {McCarthy} I.~G.,
  {Jauzac} M.,  {Li} B.,   {Schaye} J.,  2019, \mn@doi [\mnras]
  {10.1093/mnras/stz1815}, \href
  {https://ui.adsabs.harvard.edu/abs/2019MNRAS.488.3646R} {488, 3646 (R19)}

\bibitem[\protect\citeauthoryear{{Schaye} \& {Dalla Vecchia}}{{Schaye} \&
  {Dalla Vecchia}}{2008}]{Schaye2008}
{Schaye} J.,  {Dalla Vecchia} C.,  2008, \mn@doi [\mnras]
  {10.1111/j.1365-2966.2007.12639.x}, \href
  {http://adsabs.harvard.edu/abs/2008MNRAS.383.1210S} {383, 1210}

\bibitem[\protect\citeauthoryear{{Schaye} et~al.,}{{Schaye}
  et~al.}{2010}]{Schaye2010}
{Schaye} J.,  et~al., 2010, \mn@doi [\mnras]
  {10.1111/j.1365-2966.2009.16029.x}, \href
  {http://adsabs.harvard.edu/abs/2010MNRAS.402.1536S} {402, 1536}

\bibitem[\protect\citeauthoryear{{Sereno}, {Jetzer}  \& {Lubini}}{{Sereno}
  et~al.}{2010}]{2010MNRAS.403.2077S}
{Sereno} M.,  {Jetzer} P.,   {Lubini} M.,  2010, \mn@doi [\mnras]
  {10.1111/j.1365-2966.2010.16248.x}, \href
  {http://adsabs.harvard.edu/abs/2010MNRAS.403.2077S} {403, 2077}

\bibitem[\protect\citeauthoryear{{Sharon} et~al.,}{{Sharon}
  et~al.}{2020}]{2020ApJS..247...12S}
{Sharon} K.,  et~al., 2020, \mn@doi [\apjs] {10.3847/1538-4365/ab5f13}, \href
  {https://ui.adsabs.harvard.edu/abs/2020ApJS..247...12S} {247, 12}

\bibitem[\protect\citeauthoryear{{Springel}}{{Springel}}{2005}]{Springel2005}
{Springel} V.,  2005, \mn@doi [\mnras] {10.1111/j.1365-2966.2005.09655.x},
  \href {http://adsabs.harvard.edu/abs/2005MNRAS.364.1105S} {364, 1105}

\bibitem[\protect\citeauthoryear{{Wiersma}, {Schaye}  \& {Smith}}{{Wiersma}
  et~al.}{2009a}]{Wiersma2009a}
{Wiersma} R.~P.~C.,  {Schaye} J.,   {Smith} B.~D.,  2009a, \mn@doi [\mnras]
  {10.1111/j.1365-2966.2008.14191.x}, \href
  {http://adsabs.harvard.edu/abs/2009MNRAS.393...99W} {393, 99}

\bibitem[\protect\citeauthoryear{{Wiersma}, {Schaye}, {Theuns}, {Dalla Vecchia}
   \& {Tornatore}}{{Wiersma} et~al.}{2009b}]{Wiersma2009b}
{Wiersma} R.~P.~C.,  {Schaye} J.,  {Theuns} T.,  {Dalla Vecchia} C.,
  {Tornatore} L.,  2009b, \mn@doi [\mnras] {10.1111/j.1365-2966.2009.15331.x},
  \href {http://adsabs.harvard.edu/abs/2009MNRAS.399..574W} {399, 574}

\bibitem[\protect\citeauthoryear{{Wiesner} et~al.,}{{Wiesner}
  et~al.}{2012}]{2012ApJ...761....1W}
{Wiesner} M.~P.,  et~al., 2012, \mn@doi [\apj] {10.1088/0004-637X/761/1/1},
  \href {http://adsabs.harvard.edu/abs/2012ApJ...761....1W} {761, 1 (W12)}

\bibitem[\protect\citeauthoryear{{Wright} \& {Brainerd}}{{Wright} \&
  {Brainerd}}{2000}]{2000ApJ...534...34W}
{Wright} C.~O.,  {Brainerd} T.~G.,  2000, \mn@doi [\apj] {10.1086/308744},
  \href {https://ui.adsabs.harvard.edu/abs/2000ApJ...534...34W} {534, 34}

\bibitem[\protect\citeauthoryear{{van der Wel} et~al.,}{{van der Wel}
  et~al.}{2014}]{2014ApJ...792L...6V}
{van der Wel} A.,  et~al., 2014, \mn@doi [\apjl] {10.1088/2041-8205/792/1/L6},
  \href {https://ui.adsabs.harvard.edu/abs/2014ApJ...792L...6V} {792, L6}

\makeatother
\end{thebibliography}

\bsp
\label{lastpage}

\end{document}